\documentstyle[twocolumn,prl,aps]{revtex}
\begin{document}
\twocolumn[\hsize\textwidth\columnwidth\hsize\csname @twocolumnfalse\endcsname
\draft

\title{Floating of
Extended States and Localization Transition in a Weak Magnetic Field
}

\author{Kun Yang and R. N. Bhatt}
\address{Department of Electrical Engineering, Princeton University,
Princeton, NJ 08544}
\date{\today}

\maketitle

\begin{abstract}
We report results of a numerical study of non-interacting electrons moving in
a random potential in
two dimensions in the presence of a {\em weak} perpendicular magnetic
field. We study the topological
properties of the electronic eigenstates
within a tight binding model. We find that in the
weak magnetic field or strong randomness limit,
extended states float up in energy.
Further, the localization length is found to diverge at the insulator
phase boundary
with the {\em same} exponent $\nu$
as that of the isolated
lowest Landau band (high magnetic field limit).
\end{abstract}

]

Recently there has been considerable interest in the fate of delocalized
electronic states in a weak magnetic field
in two dimensions (2D)\cite{glozman,kravechenko,furneaux,shahbazyan,liu}.
In the limit of strong magnetic field, or equivalently weak randomness,
it is
believed that there exists a single critical energy within each Landau
band where the localization length of electronic states
diverges\cite{hk,bodoreview}. In contrast, one electron localization
theory\cite{gangof4} predicts
that in the absence of magnetic
field all states are localized in 2D.
Consequently, it was argued\cite{lk,laughlin} that in the limit of
weak magnetic field or strong randomness, where Landau bands merge
together, these extended
states do not disappear discontinuously but ``float
up", tending to infinite energy in the $B\rightarrow 0$ limit\cite{kramer}.
Thus, for a given electron density (and hence finite Fermi energy $E_F$),
for sufficiently low $B$
all extended states are above $E_F$
and the system becomes insulating.
This scenario
is crucial to the
global phase diagram for the quantum Hall effect
proposed by Kivelson {\em et al.}\cite{klz}
and has received strong experimental
support\cite{glozman,kravechenko,furneaux}.
Recently, however, based on numerical
calculations of localization length
on a tight binding model (TBM), Liu {\em et al.}\cite{liu}
concluded that extended states
do not float and simply
become localized as randomness increases.
This issue is more clearly posed, and its resolution well described, by
studying certain topological properties of the electronic eigenstates,
as we shall see below.

A second issue  of interest is the divergence of the localization length when
approaching the insulator-quantum Hall phase transition.
A previous numerical
study\cite{gammel} performed on a random
site TBM with a magnetic field suggested that the localization
length exponent $\nu_i\simeq 0.8$ in 2D at the localization transition
point. Besides the fact that this value is much smaller than that at the
transition between quantum Hall phases in the strong magnetic field
limit\cite{bodoreview} $\nu_H\approx 2.4$,
it violates the inequality $\nu\geq 2/d$\cite{chayes}
which is widely believed to be satisfied in
known random systems\cite{bockstedte}. To address both these issues,
a more clear-cut numerical method appears warranted.

In the presence of a magnetic field, electronic states exhibit interesting
topological properties\cite{tknn,niu,arovas,yan}.
In particular, each state can be labeled
by an integer called the Chern number, which is its boundary
condition averaged Hall conductance, in units of $e^2/h$\cite{niu,arovas}.
A state with nonzero Chern number carries Hall current and is necessarily
extended.
Thus by calculating the Chern numbers one is able to identify
extended states unambiguously on
{\em finite} size systems. This approach has proved very successful
in addressing the localization problem in the lowest Landau band\cite{huo}.
In this paper, we apply this approach to the TBM studied
by Liu {\em et al.} and also by Ando\cite{ando}. Our results
clearly support the ``floating up" picture and are consistent with
Thouless number calculations by Ando\cite{ando}. In fact,
results of Liu {\em et al.}\cite{liu}
are also consistent with ours, but our interpretation
of their results is somewhat different, as we discuss later.

We have also studied the dependence of the number and energies
of extended states
on system size. We find
just like the case of individual Landau bands, the localization length
diverges only at individual energies.
In the high field (weak randomness) limit,
the localization exponent is found to be the same as that of an isolated
lowest Landau band, $\nu_H\approx 2.4$ \cite{bodoreview}.
For strong enough randomness we find that
the localization length remains finite
throughout the band and the number of states with nonzero Chern number
goes to zero as the system size goes to infinity\cite{note1}.
Using finite size scaling, we find the largest localization length
of the system diverges as the critical randomness is reached with an
exponent $\nu_i$ which is the
same as $\nu_H$, contrary to previous suggestion\cite{gammel}
that the strong
randomness exponent may be different from that in the lowest Landau
levels. Thus our data show that $\nu$ is a universal exponent for all
spin polarized integer quantum Hall
transitions, including the ultimate one to the insulating state.

We study the TBM on a square lattice
with nearest neighbor hopping,
a uniform magnetic field and random potential,
described by the Hamiltonian:
\begin{eqnarray}
H&=&\sum_{mn}\{-t(c^\dagger_{m+1,n}c_{m,n}+c^\dagger_{m,n+1}e^{i2\pi \alpha m}
c_{m,n}+h.c.)\nonumber\\
&+&\epsilon_{m,n}c^\dagger_{m,n}c_{m,n}\},
\label{hamilt}
\end{eqnarray}
where the integers $m$ and $n$ are the $x$ and $y$ coordinates of the
lattice site in terms of lattice constant,
$c_{m,n}$ is the fermion operator on that site, $t$ is the
hopping matrix element which we set as the unit of energy from now on,
and $\epsilon$ is the random potential ranging {\em uniformly} from $-W$ to $W$
(as in the Anderson model\cite{anderson}).
$\alpha$ is the amount of
magnetic flux per plaquette
in units of the flux quantum $hc/e$.
The Landau gauge ${\bf A}=(0, Bx, 0)$ is used in Eq. (\ref{hamilt}).
Here we concentrate
on the case $\alpha=1/N_f$, where $N_f$ is an integer. In this case, we have
$N_f$ Landau subbands in the absence of random potential, and the lowest
energy
subbands map onto the lowest Landau levels in the
limit $N_f\rightarrow\infty$,
which is the continuum limit.

The Hall conductance of an individual eigenstate $|m\rangle$ can be
obtained easily using the Kubo formula:\cite{yan}
\begin{eqnarray*}
\nonumber
\sigma_{xy}^{m}={ie^2\hbar\over A}\sum_{n\ne m}{\langle m|v_y|n\rangle
\langle n|v_x|m\rangle-\langle m|v_x|n\rangle\langle n|v_y|m\rangle\over
(E_n-E_m)^2},\nonumber
\end{eqnarray*}
where $A$ is the area of the system, $v_x$ and $v_y$ are the velocity operators
in the $x$ and $y$ directions respectively. For a finite system
with the geometry of a parallelogram with periodic boundary conditions
(torus geometry), $\sigma_{xy}^m$ depends on the two boundary
condition phases $\phi_1$ and $\phi_2$. As shown by Niu {\em et al.},
the boundary condition averaged Hall conductance takes the
form\cite{niu}
\begin{equation}
\langle\sigma_{xy}^m\rangle={1\over 4\pi^2}\int{d\phi_1d\phi_2\sigma_{xy}^{m}
(\phi_1,\phi_2)}=C(m)e^2/h,
\end{equation}
where $C(m)$ is an integer called the Chern number of the state $|m\rangle$.
States with nonzero Chern numbers carry Hall
current and are necessarily
extended states\cite{arovas,huo}. Thus by numerically diagonalizing
the Hamiltonian on a grid of $\phi_1$ and $\phi_2$, and calculating the
Chern numbers of states of finite size systems by converting the integral in
(2) to a sum over grid points, we are able to identify
extended states unambiguously.

We have studied systems of square geometry
with various size (from $3\times 3$
to $15\times 15$), strength of randomness ($W$) and
magnetic field (equivalently, $N_f$). The number
of samples explored for a given $W$ range from
$2000$ to $30$ depending on system size. Most of our data were taken for
$N_f=3$. We do not, however, see any qualitative difference in
behavior of the extended states, for systems
with $N_f$ as large as 13. Hence we believe our results
with $N_f=3$ are generic and apply to the continuum limit $N_f\rightarrow
\infty$.

Fig. \ref{fig1} shows the density of states [$\rho(E)$]
and density of extended states
with nonzero Chern numbers [$\rho_c(E)$],
for two different strength of randomness for $1/3$ flux quantum per plaquette
($N_f=3$) on a square of lattice size $9\times 9$.
For weak enough randomness ($W=1.0$),
the three Landau subbands are broadened by randomness, but are still well
separated.
We see there are extended states in all subbands, with their densities peaked
essentially at the center of each subband.
This is consistent with the
previous study on individual Landau bands\cite{huo}.
As randomness increases, the subbands further broaden and start to merge,
as is seen for $W=2.5$.
In this case there are still three
prominent peaks in $\rho(E)$ (we call them $E_1$, $E_2$ and $E_3$
respectively),
which are (loosely) identified as centers of Landau subbands.
$\rho_c(E)$, however, now looks very different:
most of the
extended states are near the center of the entire band ($E_2$) and there is
no peak in $\rho_c(E)$
at $E_1$ or $E_3$, which are the centers of
Landau subbands. There are nontrivial features in $\rho_c(E)$ which we
discuss below, but it is clear from Fig. \ref{fig1} that as the three
subbands start to merge, the extended states in the lower and upper
subbands move away from the centers of the subbands ($E_1$ and $E_3$)
toward center of the
band ($E_2$). This behavior is also seen in systems of
$N_f$ as large as 13. We hence
believe in the limit $N_f\rightarrow\infty$
(which can be mapped onto the continuum model), the extended states
in the lowest subbands (which becomes Landau levels) float up toward the center
of the band (which is at infinitely high energy
relative to them in the continuum model).
This provides unambiguous support for the floating up picture predicted
theoretically\cite{lk} and seen experimentally\cite{glozman}.

The fact that the extended states in the lower and upper subbands float
toward the center of the band as randomness increases may be understood
in the following manner.
In finite size systems, the Chern number of a state can change only when
it becomes degenerate with a another state under certain boundary conditions.
This can be shown to occur only by tuning three parameters, including
the two boundary condition angles plus the parameter characterizing the
random potential. If such a degeneracy were to occur, the Chern numbers
of the two states involved may change but their sum is conserved.
Randomness tends to localize all states and annihilate the nonzero Chern
numbers carried by the extended states. Thus states with nonzero Chern numbers
of opposite signs ``attract" each other and tend to move close in energy
as randomness increases.
It is believed that in the thermodynamic limit
the localization length diverges and true current carrying
(extended) states exist only
at individual critical energies.
(We
will provide numerical evidence for this belief below.)
Each such critical energy
is characterized by its total Chern number which is {\em invariant} as
randomness varies, unless merging between critical energies occur.
For exactly the same reason, critical energies with total Chern numbers of
opposite sign also ``attract" each other as randomness increases.
In the case of $N_f=3$ systems, the total Chern numbers for the
three subbands are 1, -2 and 1 respectively.
Due to the ``attraction", we expect that as randomness is turned
on, the extended states in the central subband with
total Chern number $-2$ splits into two critical energies
with total Chern number $-1$ each (by symmetry)
and move toward the two band edges as randomness is increased
further. Concurrently,
the two critical energies
of the upper and lower subbands with total Chern number +1 move
away from the center of the subbands
toward the center of the band.
This is precisely what is seen in the $\rho_c(E)$ at $W=2.5$:
There is a small dip at the
center of the band indicating the splitting of the central critical energy;
further, there are
two less pronounced peaks from the two edge subbands, whose positions have
clearly moved away from the corresponding peaks of $\rho(E)$.

Fig. \ref{fig2} depicts
the number of states with nonzero Chern number
$N_c\equiv\int_{-\infty}^{\infty}{\rho_c(E)dE}$
versus the system size $N_s$ (number of
sites), for different values of disorder $W$, for
$N_f=3$, on a double logarithmic plot. We find the plot is essentially
linear for small $W$ up to $W\approx3.0$,
with slope
$y=0.79\pm 0.01$ which is relatively independent of
$W$\cite{linearnote}, indicating that
$N_c\sim (N_s)^y$ in this region.
This power law behavior
is exactly what is expected\cite{yan,huo} where there
are individual critical energies $E_c^i$ in the vicinity of which
the localization length
diverges with a power law of the form $\xi(E)\sim |E-E_c|^{-\nu}$.
In a finite system with linear size $L_s=\sqrt{N_s}$,
states with $\xi(E)>L_s$ look extended.
The number of such states goes like
$N_c\sim N_s\rho(E_c)L_s^{-1/\nu}\sim N_s^{1-1/2\nu}$,
thus
$y=1-1/2\nu$. This gives
$\nu=2.4\pm 0.1$, in agreement with the $\nu_H$ for
lowest Landau band\cite{bodoreview,huo}.
This suggests that $\nu$ is a universal
exponent in all spin-polarized integer quantum Hall
transitions.

For larger $W$, the dependence of $N_c$ on $N_s$ deviates from a power
law
and bends down as $N_s$ increases. This indicates that in this
regime the two critical energies with total Chern number -1 have merged with
the other two with Chern number +1;
all extended states have disappeared and $\xi$ is finite throughout the
band. For strong enough randomness and large $N_s$, $N_c$ {\em decreases}
as $N_s$ increases; thus in the localized regime
the average number of extended states
per sample goes to zero in the thermodynamic limit.
{}From the shape of the density of extended states and scaling of data
we determine the critical randomness to be $W_c\approx 2.9\pm 0.1$.
For $W$ greater than
but close to $W_c$, and large sizes $N_s$,
$N_c$ is expected to take the scaling
form:
$N_c\sim N_s^y \tilde{F}(L_s/\xi_m)\sim N_s^y F(N_s^{1/(2\nu_i)}(W-W_c))$,
where $\xi_m$ is the largest localization length in the system that diverges
as $W_c$ is approached with exponent $\nu_i$.
The best scaling is achieved with $\nu_i\approx 2.3$,
assuming $W_c=2.9$, and Fig. \ref{fig3} shows the scaling function $F$.
Taking into uncertainty in $W_c$ we estimate
$\nu_i\approx 2.3\pm 0.3$.
This suggests that the localization length exponents are the same in both
the localized and extended regimes, in contrast to a previous suggestion that
they may be different\cite{gammel}.
The increasing negative slope of the scaling curve suggests that $N_c$
goes to zero faster than any power law as $N_s$
increases at large $N_s$.

We emphasize that the existence of this localized regime in the TBM
is due to the facts that there exist critical energies with negative
Chern numbers and the total Chern number of the system
is zero. In the continuum however,
the total Chern number of the critical energy
of each Landau band is one,
and
there is {\em no} critical energy with negative Chern number
at finite energy. Hence the extended states at these
 critical energies
cannot annihilate their
Chern numbers and become all
localized as the randomness increases. They only float up and ``disappear"
at infinite energy. This becomes clear as one views the continuum system as
the $N_f\rightarrow\infty$ limit of the TBM. In the TBM,
the natural energy scale is hopping $t$ (set to be 1 previously),
and the zero point of energy is the center of the
band. In the continuum however, the energy scale is Landau level
spacing $\hbar\omega_c$, and the zero point of energy is
determined by identifying the center of the lowest energy band with
energy $\hbar\omega_c/2$. In terms of TBM
parameters we have $\hbar\omega_c=4t/N_f$.
Based on our data up to $N_f=13$ we conclude that
the critical randomness is
almost $N_f$ independent
and is about $W_c\approx 3t$, in agreement with Ando\cite{ando}. The energy
at which the final merging and disappearance of critical energies measured
from the {\em bottom} of the band is found to be
of order $O(W_c)$, which is the only energy scale of the TBM at
criticality.
Therefore the number of Landau subbands below the lowest
critical energy before it finally disappears is of order
$O(W_c/\hbar\omega_c)\propto N_f$.
We hence conclude in the continuum limit ($\hbar\omega_c$ finite,
$N_f\rightarrow\infty$) the critical randomness strength
($W_c\propto N_f\hbar\omega_c$) is
infinite, and extended states all float up to infinite
energy in the strong randomness (or weak magnetic field) limit.

Liu {\em at al.}\cite{liu}
interpret the localization transition in the TBM
as indication of disappearance of extended states in the continuum. As
discussed above, the continuum limit of the TBM is subtle.
They also find
the energies of the extended states do not shift much relative to the
{\em center} of the band ($E=0$) and interpret it as evidence against floating.
Our results for $N_f=3$ is consistent with very little shift of $E_1$ and $E_3$
relative to $E_2$.
However, critical energies clearly float away from peaks of $\rho(E)$
(which are roughly at the centers of Landau subbands),
and more so relative to the {\em bottom} of the band.
This is because
as randomness increases, both the bottom of the band and peaks in $\rho(E)$
move {\em downward}. We believe this
{\em relative} movement is clear indication
of floating of extended states which survives the continuum limit.

In summary, we have found unambiguous numerical evidence, using a tight
binding model (TBM) and considering the passage to the continuum limit,
which indicates that
extended states float up in energy toward infinity in the weak
magnetic field limit in 2D in the continuum.
For the TBM, this can be heuristically understood in terms of an
``attraction" between states with opposite
Chern numbers. There is a critical randomness strength in the TBM at
which all states become localized. The localization length diverges with
the same exponent as that of the isolated lowest Landau band (high field
limit) when approaching this critical point.

We thank M. Guo, F.D.M. Haldane,
B. Huckestein and D. Shahar for helpful conversations.
This work was supported by NSF grants DMR-9224077,
DMR-9400362 and CDA-9121709. One of us (RNB) was also supported in part by a
Guggenheim fellowship.

\begin{figure}
\caption{Ensemble averaged density of states $\rho(E)$ and
density of extended states $\rho_c(E)$
for two values of randomness $W$, for systems of size
$9\times 9$.}
\label{fig1}
\end{figure}

\begin{figure}
\caption{Number of extended states $N_c$ versus system size $N_s$ for
various $W$ on a double logarithmic scale.
The solid line with slope $y=0.79$
is a linear fit to the data for $W=3.0$.
}
\label{fig2}
\end{figure}

\begin{figure}
\caption{The scaling function $F(N_s^{1/(2\nu_i)}(W-W_c))$.}
\label{fig3}
\end{figure}

\begin{references}
\bibitem{note1} This localized regime is special to
this one band model.
Continuum systems in a magnetic field
always have some extended states. See below.
\bibitem{linearnote} Actually for $W\leq 2.5$ the curves seem to
deviate slightly upwards
from linear behavior. This may be
due to the existence of close lying critical energies between which the
localization length is larger than the system sizes we have studied, so the
system appears to have a narrow band of extended states instead
of isolated critical energies.
\end{references}
\end{document}